\documentclass[12pt,a4paper,fleqn]{article}

\usepackage[DIV12]{typearea}
\usepackage{amsmath,amsfonts,amssymb}
\usepackage{graphicx}
\usepackage{cite}
\usepackage{mathrsfs}
\usepackage{bbm}
\usepackage{dsfont}
\usepackage{units}
\usepackage{subfig}
\usepackage{lscape}

\newcommand{\Figref}[1]{figure~\ref{#1}}

\DeclareMathOperator{\tr}{Tr}

\newcommand{\D}{\text{d}}

\newcommand{\eVdist}{\kern-0.06em}

\newcommand{\SU}[1]{\ensuremath{\mathrm{SU}(#1)}}

\newcommand{\Y}{\ensuremath{\boldsymbol{Y}}}
\newcommand{\Yu}{\ensuremath{\boldsymbol{Y}_{\!\!u}}}
\newcommand{\Yd}{\ensuremath{\boldsymbol{Y}_{\!\!d}}}
\newcommand{\Ye}{\ensuremath{\boldsymbol{Y}_{\!\!e}}}

\allowdisplaybreaks[1]

\begin{document}

\begin{titlepage}

\begin{flushright}
TUM-HEP 687/08
\end{flushright}

\vspace*{1.0cm}

\begin{center}
{\Large\bf 
Running minimal flavor violation
}

\vspace{1cm}

\textbf{
Paride Paradisi\footnote[1]{Email: \texttt{Paride.Paradisi@uv.es}},
Michael Ratz\footnote[2]{Email: \texttt{mratz@ph.tum.de}},
Roland Schieren\footnote[3]{Email: \texttt{Roland.Schieren@ph.tum.de}},
Cristoforo Simonetto\footnote[4]{Email: \texttt{Cristoforo.Simonetto@ph.tum.de}}
}
\\[5mm]
\textit{\small
Physik-Department T30/T31, Technische Universit\"at M\"unchen, \\
James-Franck-Stra\ss e, 85748 Garching, Germany
}
\end{center}

\vspace{1cm}

\begin{abstract}
We consider the flavor structure of the minimal supersymmetric standard model
(MSSM) in the framework of `minimal flavor violation' (MFV). We show that, if
one imposes the MFV structure at some scale, to a good accuracy the MFV
decomposition works at all other scales. That is, quantum effects can be
described by running coefficients of the MFV decomposition. We find that the
coefficients get driven to non-trivial fixed points. 

\end{abstract}

\end{titlepage}

\newpage

\section{Introduction}

Despite of its great phenomenological success, the standard model (SM) is
certainly not completely satisfactory from a theoretical point of view. Certain
aspects of the SM  hint at unified structures: the gauge interactions and the
quantum numbers of the fundamental fermions fit nicely into the framework of a
grand unified theories (GUTs).  As is well known, GUTs seem to require
low-energy supersymmetry, as this most allows for the compelling scenarios of
gauge unification. This leads to the picture of the so-called `SUSY desert',
i.e.\ between the TeV scale and the GUT scale no new physics appears.
On the other hand, attempts to find a simple explanation of the SM flavor
structure have not yet been as successful as one could have hoped.

In this study we consider the minimal supersymmetric extension of the SM (MSSM), where the flavor structure is
particularly rich because of the various additional soft terms.
As is well known, the flavor parameters are tightly constrained by
phenomenology, leading to what is usually called the supersymmetric flavor
problems. These problems may be viewed as evidence against low-energy
supersymmetry. Adopting a more optimistic point of view, one could say that the
non-observation of certain flavor transitions enforces a rather special form of
soft terms, so that one can gain additional insights on the origin of flavor by
studying superpartner interactions.

An efficient way to ameliorate (or even avoid) the supersymmetric flavor
problems is to assume that the (soft) masses of the squarks and sleptons, i.e.\
the scalar superpartners of SM quarks and leptons, are close to a unit matrix.
In this case the super-GIM mechanism is at work \cite{Dimopoulos:1981zb}, i.e.\
unobserved flavor transitions are strongly suppressed. It has been rather
popular to assume that soft masses are proportional to the unit masses at a high
scale, such as the GUT scale, and all deviations come from radiative
corrections, induced by the Yukawa couplings. However, one might argue that this
assumption lacks a fundamental motivation. 

In this note, we consider a slightly modified setting in which this strong
assumption gets somewhat relaxed. We shall assume that at the GUT scale
the scalar soft mass squareds receive corrections that are proportional to
$\Y^\dagger \Y$ where $\Y$ denotes a Yukawa coupling matrix. In other words, we
study the implications of an ansatz which is known as `minimal flavor violation'
(MFV) \cite{Chivukula:1987py,Buras:2000dm,D'Ambrosio:2002ex} at the GUT scale. 

\section{A short review of the MFV ansatz}

As is well known, the MFV ansatz is motivated as follows: in the limit of
vanishing Yukawa couplings the MSSM enjoys an enhanced (classical) symmetry,
\begin{equation}\label{eq:Gflavor}
 G_\mathrm{flavor}~=~
\SU3_u \times \SU3_d \times \SU3_Q \times \SU3_e \times \SU3_L\;.
\end{equation}
One might then view the Yukawas as vacuum expectation values (vevs) of 
`spurion' fields. If these spurions are the only source of flavor violation,
this implies that any operator not respecting $G_\mathrm{flavor}$ has to be
proportional to the spurions, i.e.\ to the Yukawas. This then leads to the
following expansion of soft supersymmetry breaking operators \cite{D'Ambrosio:2002ex}:
\begin{subequations}\label{eq:softmassdecomp}
\begin{eqnarray}
  \boldsymbol{m}_Q^2 &= & \alpha_1\,\mathds{1} + \beta_1 \,\Yu^\dagger \Yu + \beta_2\, \Yd^\dagger \Yd 
 + \beta_3\, \Yd^\dagger \Yd\, \Yu^\dagger \Yu + \beta_3 \,\Yu^\dagger \Yu\, 
 \Yd^\dagger \Yd\;, \\
 \boldsymbol{m}_u^2 &= & \alpha_2\,\mathds{1} + \beta_5\, \Yu \Yu^\dagger  \;,\\
 \boldsymbol{m}_d^2 &= & \alpha_3\,\mathds{1} + \beta_6\, \Yd \Yd^\dagger  \;,\\
 \boldsymbol{A}_u &= &  \alpha_4\, \Yu + \beta_7\, \Yu\, \Yd^\dagger \Yd  \;,\\
 \boldsymbol{A}_d &= & \alpha_5\, \Yd + \beta_8\, \Yd \,\Yu^\dagger \Yu  \;,\\
 \boldsymbol{A}_e &= &\alpha_e\, \Ye \;.
\end{eqnarray}
\end{subequations}
Higher order terms in this expansion can be neglected due to the Yukawa
hierarchies. Note that our notation is slightly different from the one used in
\cite{D'Ambrosio:2002ex} in that our coefficients $\alpha_i$ and $\beta_i$ carry
mass dimension. This is done in order to simplify the expressions to be
presented below. In our notation of Yukawa couplings and scalar soft mass
squareds we follow \cite{Martin:1993zk}.

As discussed above, the MFV ansatz offers a natural way to avoid  unobserved
large effects in flavor physics. However, we would like to  stress here that
small departures from complete flavor blindness of the soft terms can
still provide interesting effects in low energy processes. In particular, the
$\beta_1 \,\Yu^\dagger \Yu$ term in $\boldsymbol{m}_Q^2$ induces Flavor
Changing Neutral Currents (FCNC) phenomena, such as $B\to X_s\gamma$, through a
loop exchange of gluinos and squarks. For $\mathcal{O}(1)$ values of the MFV
parameter $\beta_1$, sizable or even dangerous contributions to flavor physics
observables can be expected, depending on the absolute soft SUSY scale.
Additionally, from a phenomenological side, it is well known that both FCNC
transitions and the prediction for the lightest Higgs boson mass are highly
sensitive to the $\boldsymbol{A}_u$ term in the stop sector thus, the  MFV
modifications to $\boldsymbol{A}_u$ can play, in principle, a relevant role. In
this respect, one might expect that low-energy observables can still
represent a useful tool to test and/or constrain the MFV parameters at the
high scale. In particular, one would expect to find departures from the
predictions of mSUGRA models where a completely flavor blind scenario is
realized at the high scale. However, as we will discuss in the next sections,
this is not the case.

\section{MFV decomposition and renormalization effects}

\subsection{Scale-independent validity of the MFV ansatz\label{ch:validity}}

Usually the MFV ansatz is imposed at a scale close to the electroweak scale. As
mentioned in the introduction, we will be interested in the situation where it
is imposed at the GUT scale. Since the spurion argument does not
imply a preferred scale, one might expect that, if the MFV decomposition
applies at one renormalization scale, it will apply at different scales as
well. That is, renormalization effects will modify the values of the
coefficients, $\alpha_i$ and $\beta_i$, but not the validity of the ansatz. 

We have checked explicitly that this is the case: we start with soft terms
complying with the decomposition \eqref{eq:softmassdecomp} at the GUT scale
and run them down to the SUSY scale, i.e.\ solve the corresponding
renormalization group equations (RGEs). Then we successfully fit the low energy
soft masses by the decomposition \eqref{eq:softmassdecomp}, i.e.\ when inserting the
Yukawa matrices at the low scale we find values of the MFV parameters $\alpha_i$
and $\beta_i$ such that the mass matrices are reproduced with a high
accuracy. The details of our numerical studies are deferred to
appendix~\ref{app:MFVcheck}.

\subsection{RGEs for the MFV parameters \label{ch:RGEs_MFV}}

Having seen that the running of the soft masses can described in terms of scale
dependent MFV coefficients $\alpha_i$ and $\beta_i$ we now study the behavior of
these coefficients under the renormalization group.
We calculate the RGEs for the $\alpha_i$ and $\beta_i$ by inserting
\eqref{eq:softmassdecomp} in the one-loop RGEs for the soft-masses and the
trilinear couplings (cf.\ \cite{Martin:1993zk}). Note that there are two sources
for the running of the MFV coefficients: first, the soft terms run, and second,
the Yukawa matrices, to which we match the soft terms, also depend on the
renormalization scale. Neglecting the Yukawa couplings of the first and second
generation, the results read
\begin{subequations}\label{eq:MFVRGEs}
\begin{eqnarray}
 16\pi^2 \frac{\D \alpha_1}{\D t} 
 &=& -\frac{32}{3} g_3^2 |M_3|^2 - 6g_2^2|M_2|^2 - \frac{2}{15} g_1^2|M_1|^2 +
 \frac{1}{5}g_1^2 S\;, \\
 16\pi^2 \frac{\D \alpha_2}{\D t} 
 &=& -\frac{32}{3} g_3^2 |M_3|^2 - \frac{32}{15} g_1^2|M_1|^2 - \frac{4}{5}g_1^2
 S\;, \\
 16\pi^2 \frac{\D \alpha_3}{\D t} 
 &=& -\frac{32}{3} g_3^2 |M_3|^2 - \frac{8}{15} g_1^2|M_1|^2 + \frac{2}{5}g_1^2
 S\;, \\
 16\pi^2 \frac{\D \alpha_4}{\D t} 
 &=& 12 \alpha_4 y_t^2 + 10 \beta_7y_t^2 y_b^2 + 2 \beta_8
 y_t^2y_b^2 +  \frac{32}{3}g_3^2 M_3 + 6 g_2^2 M_2 + \frac{26}{15}
 g_1^2 M_1\;,  \\
 16\pi^2 \frac{\D \alpha_5}{\D t} 
 &=& 12 \alpha_5 y_b^2 +
 10 \beta_8y_t^2 y_b^2 + 2\beta_7 y_t^2y_b^2 + 
 \frac{32}{3}g_3^2 M_3 + 6 g_2^2 M_2 + \frac{14}{15} g_1^2 M_1 + 2 \alpha_e
 y_\tau^2\;, \nonumber\\
 & &\\
 16\pi^2 \frac{\D \beta_1}{\D t} 
 &=& 2 m_{H_u}^2 + 2 \alpha_4^2 + 2\beta_8^2y_t^2 y_b^2 +  2\alpha_1
 + 2\alpha_2 \nonumber\\
 & & {}- 10\beta_1y_t^2 +2 \beta_5y_t^2 + \beta_1 \left(
 \frac{32}{3}g_3^2 + 6 g_2^2 + \frac{26}{15}g_1^2 \right)\;, \\
 16\pi^2 \frac{\D \beta_2}{\D t} 
 &=& 2 m_{H_d}^2 + 2\alpha_5^2 + 2
 \beta_7^2y_t^2 y_b^2 +  2\alpha_1 + 2\alpha_3 -
 10\beta_2y_b^2 - 2\beta_2y_\tau^2 + 2\beta_6y_b^2
 \nonumber\\
 &&{}+ \beta_2 \left( \frac{32}{3}g_3^2 + 6 g_2^2 + \frac{14}{15}g_1^2
 \right)\;, \\
 16\pi^2 \frac{\D \beta_3}{\D t} 
 &=& 2 \alpha_4\beta_7 +2 \alpha_5\beta_8 -12\beta_3y_t^2 -
 12\beta_3y_b^2 - 2\beta_3y_\tau^2  + \beta_3
 \left(\frac{64}{3}g_3^2 + 12g_2^2 + \frac{8}{3}g_1^2 \right)\;,   \nonumber\\
 & & \\
 16\pi^2 \frac{\D \beta_5}{\D t} 
 &=& 4 m_{H_u}^2 +4 \left( \alpha_4 + \beta_7y_b^2 \right)^2 +  4\alpha_1
 + 4\alpha_2 + 4 \beta_1y_t^2 + 4\beta_2y_b^2 +
 8\beta_3y_t^2y_b^2 \nonumber\\
 & &{}   + \beta_5 \left( -8y_t^2-2y_b^2 +\frac{32}{3}g_3^2 + 6 g_2^2 + \frac{26}{15}g_1^2 \right)\;,  
 \\
 16\pi^2 \frac{\D \beta_6}{\D t} 
 &=& 4 m_{H_d}^2 + 4\left( \alpha_5 + \beta_8y_t^2\right)^2 +  4\alpha_1 + 4\alpha_3 + 4\beta_1y_t^2 + 4\beta_2y_b^2 + 8\beta_3y_t^2y_b^2 
 \nonumber\\
 & & {} + \beta_6\left( - 2y_t^2 - 8y_b^2-2y_\tau^2 +
 \frac{32}{3}g_3^2 + 6g_2^2 + \frac{14}{15}g_1^2  \right)\;, \\
 16\pi^2   \frac{\D \beta_7}{\D t} 
 &=& 2\alpha_5 + \beta_7\left( -12y_b^2 - 2y_\tau^2 +
 \frac{32}{3}g_3^2 + 6g_2^2 + \frac{14}{15}g_1^2 \right)\;, \\
 16\pi^2   \frac{\D \beta_8}{\D t} 
 &=& 2\alpha_4 + \beta_8\left( -12y_t^2 +
 \frac{32}{3}g_3^2 + 6g_2^2 + \frac{26}{15}g_1^2 \right)\;.
 \end{eqnarray}
\end{subequations}
Here $\D/\D t$ denotes the logarithmic derivative w.r.t.\ the
renormalization scale, $g_1$, $g_2$, $g_3$ are the gauge couplings,
$M_1$, $M_2$, $M_3$ the gaugino masses,  $y_t$,
$y_b$, $y_\tau$ the third family Yukawa couplings, $m_{H_u}$, $m_{H_d}$
the Higgs soft mass terms. We have further defined
\begin{eqnarray}
 S &=& 
 m_{H_u}^2 - m_{H_d}^2 + \mathrm{Tr}\bigg[ \alpha_1\mathds{1} + \beta_1 \Yu^\dagger \Yu + \beta_2 \Yd^\dagger \Yd + 2 \beta_3 \Yd^\dagger \Yd \Yu^\dagger \Yu 
 \nonumber\\
 && {}- 2 \alpha_2\mathds{1} - 2 \beta_5 \Yu \Yu^\dagger  
  + \alpha_3\mathds{1} + \beta_6 \Yd \Yd^\dagger - \boldsymbol{m}_L^2 + \boldsymbol{m}_e^2 \bigg] 
\end{eqnarray}
with $\boldsymbol{m}_L^2$ and $\boldsymbol{m}_e^2$ denoting the $3\times3$ mass
matrices for the charged lepton doublets and singlets, respectively.

\subsection{Approximations of low-energy MFV coefficients}

We derive approximate relations between the values at the GUT scale and the low scale.
Here we assume mSUGRA inspired initial conditions~(for details see
equation~\eqref{eq:universal} in appendix~\ref{app:Approximations})
and allow for one non-zero $\beta_i$ while the others are set to zero. 
The value of $\tan\beta$ is fixed to 10. The formulae are obtained by varying the initial
values of $m_{\nicefrac{1}{2}},m_0,A,\beta_i$, running them down to the low
scale and fitting a linear combination of the parameters to the obtained points
in parameter-space.  Details and the results are shown in appendix~\ref{app:Approximations}.

\subsection{Fixed points in the evolution of the MFV coefficients}

Let us now come to the discussion of the relation between the boundary values
for the MFV coefficients at the high scale and the values they attain at the low
scale. 
A crucial feature of the low-energy values of the MFV coefficients $\beta_i$ is
that they are rather insensitive to their GUT boundary values. 
It is, of course,
well known that the soft masses tend to get aligned due to the renormalization
group evolution
\cite{Brignole:1993dj,Choudhury:1994pn,Brax:1995up,Chankowski:2005jh}. Our
results make this statement more precise. There is an on-going competition
between the alignment process, triggered mainly by the positive gluino
contributions, and misalignment process, driven by negative effects proportional
to the Yukawa matrices. 
These effects are so strong that the memory to the initial conditions gets
almost wiped out, at least as long as the ratio between scalar and gaugino
masses at the high scale is not too large.

\begin{figure}
 \centering
 \subfloat[SPS1a]{\includegraphics{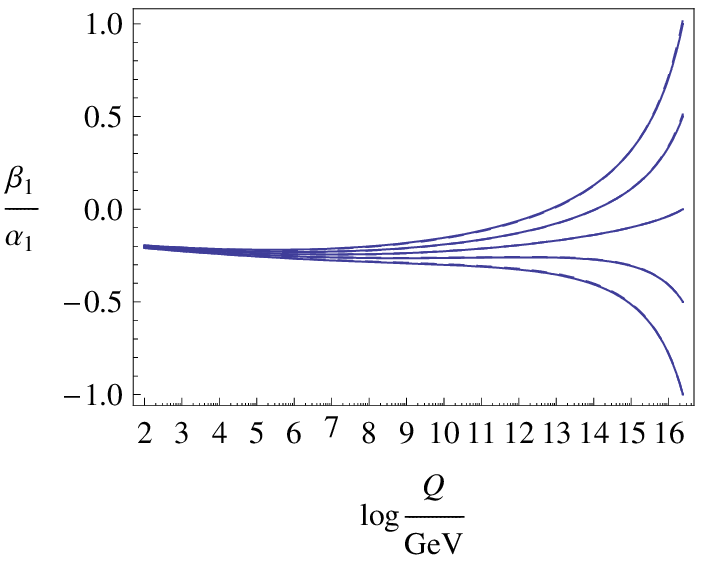}}
 \quad 
 \subfloat[SPS1b]{\includegraphics{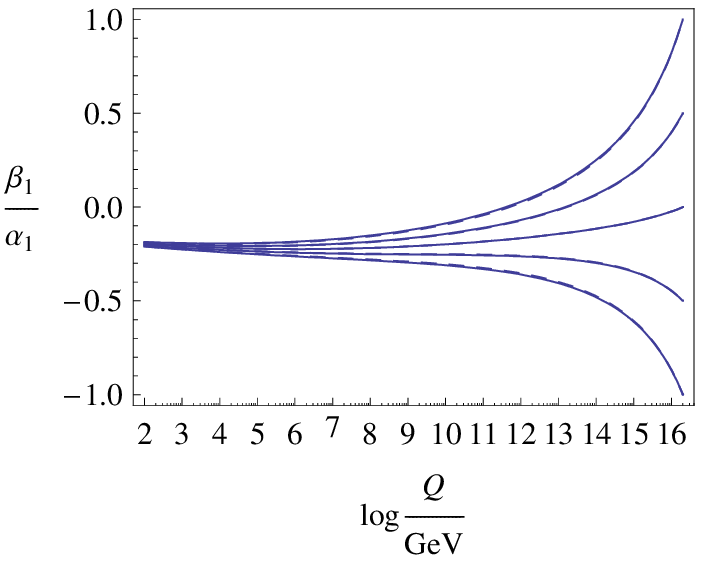}}
 \\[0.3cm]
 \subfloat[SPS2]{\includegraphics{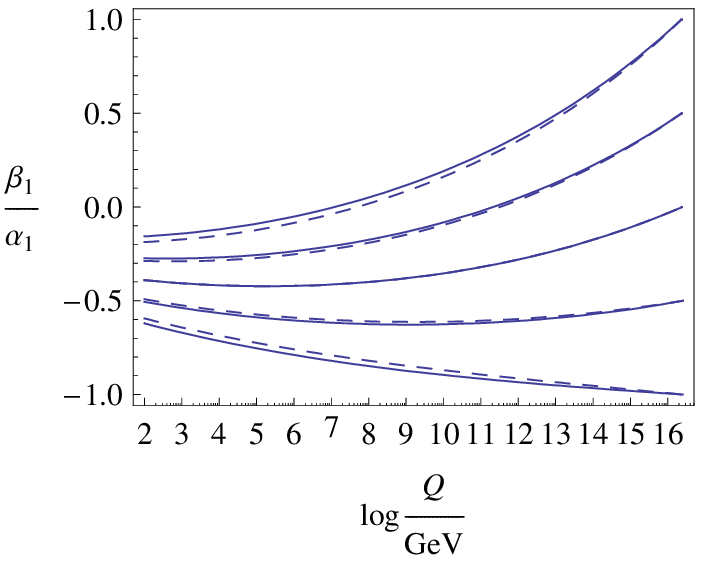}}
 \quad 
 \subfloat[SPS3]{\includegraphics{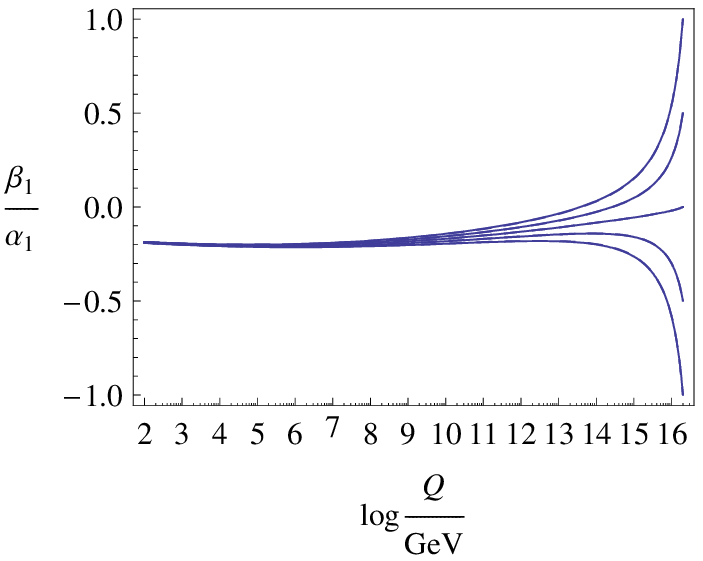}}
 \\[0.3cm]
 \subfloat[SPS4]{\includegraphics{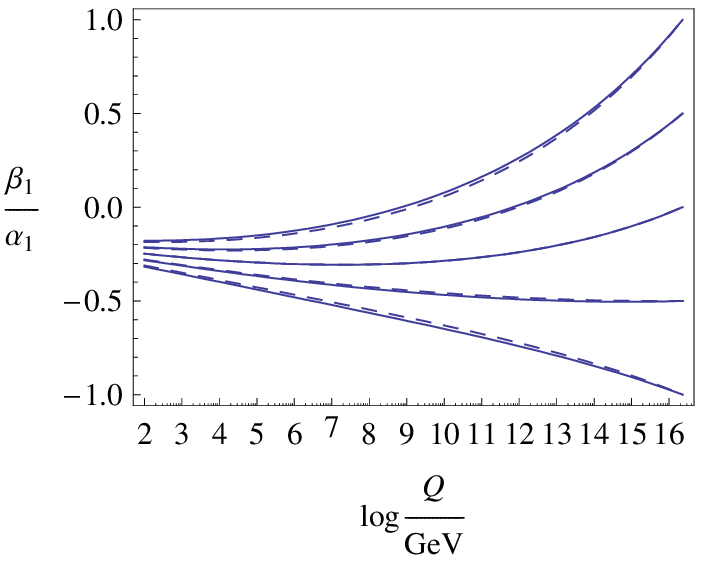}}
 \quad 
 \subfloat[SPS5]{\includegraphics{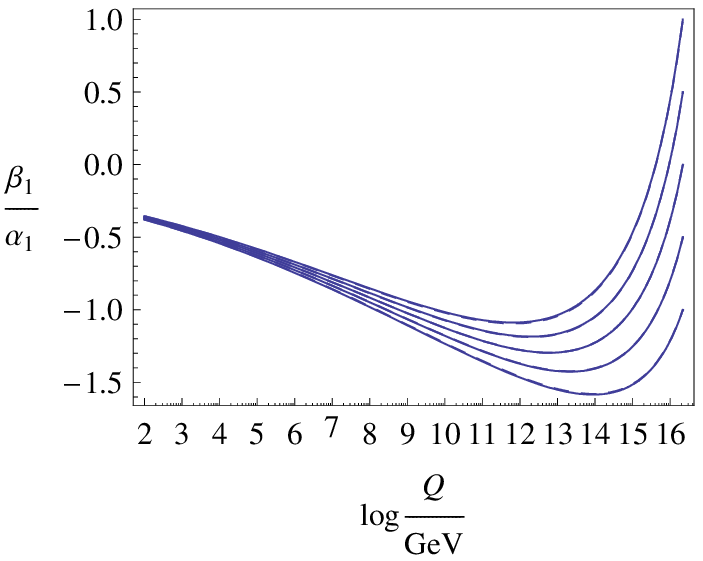}}
 \caption{The running of $\frac{\beta_1}{\alpha_1}$. For the solid curve
  only $\beta_1$ is non-zero at the high scale while for the dashed curve all
  $\beta_i$ are switched on. }
 \label{fig:b_1/a_1}
\end{figure}

\begin{figure}
 \centering
 \subfloat[SPS1a]{\includegraphics{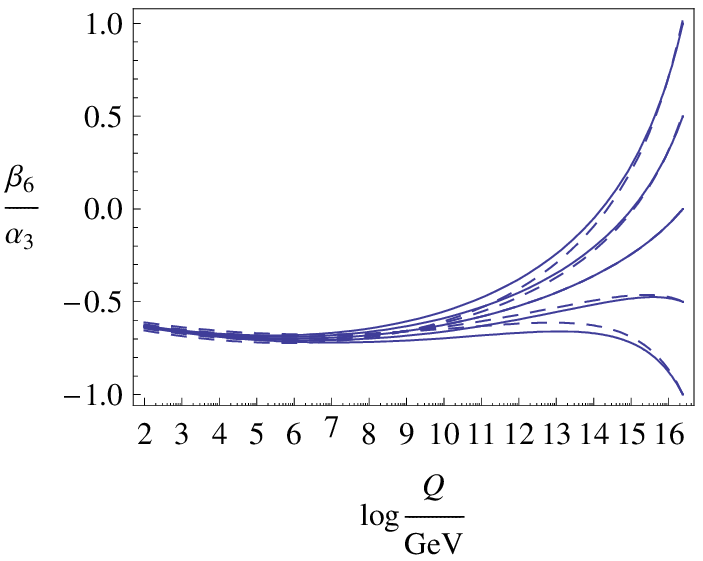}} 
 \quad
 \subfloat[SPS1b]{\includegraphics{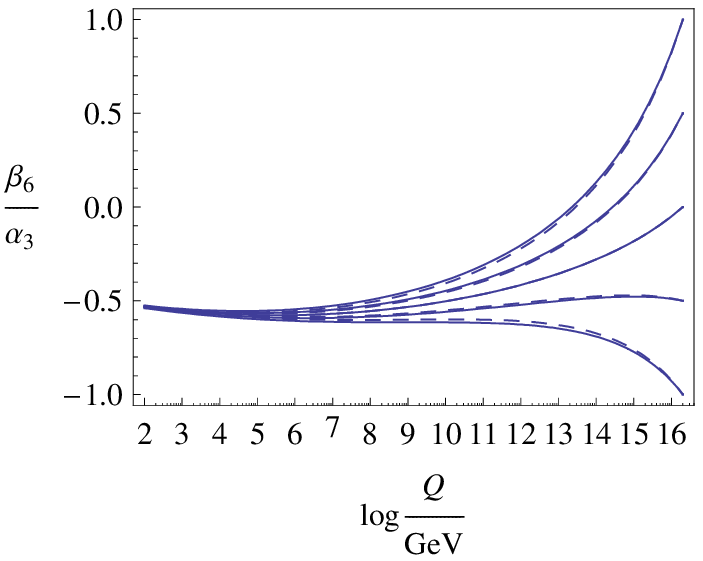}}
 \\[0.3cm]
 \subfloat[SPS2]{\includegraphics{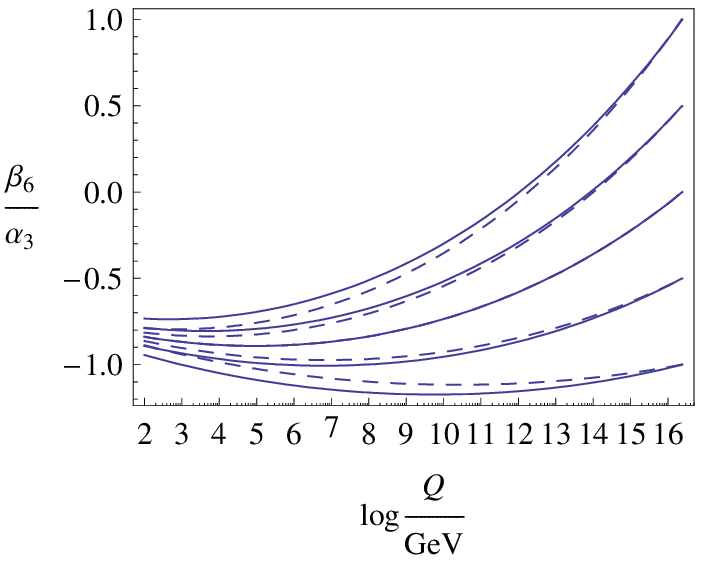}}
 \quad
 \subfloat[SPS3]{\includegraphics{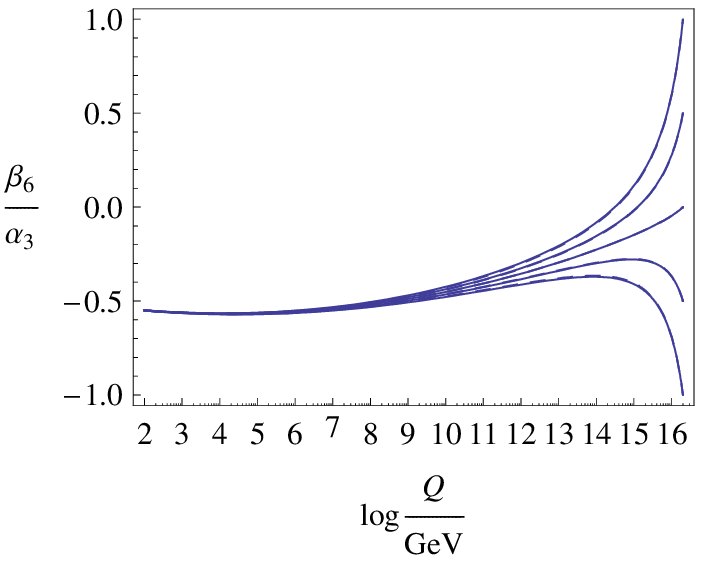}}
 \\[0.3cm]
 \subfloat[SPS4]{\includegraphics{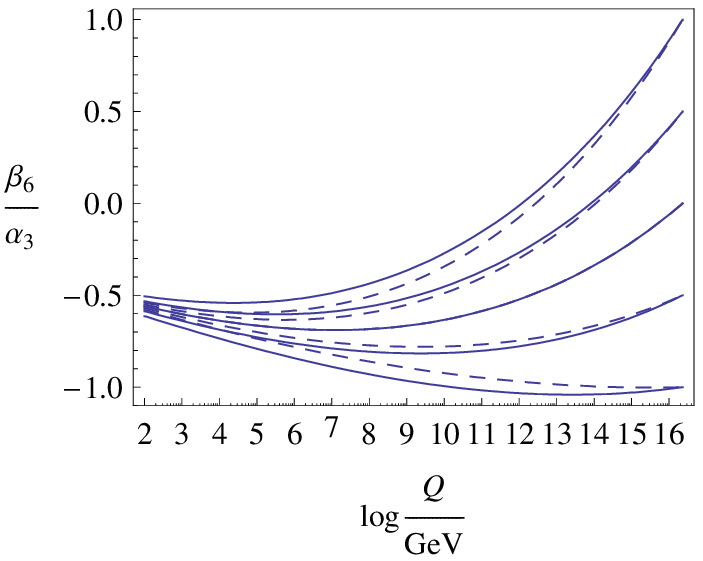}}
 \quad
 \subfloat[SPS5]{\includegraphics{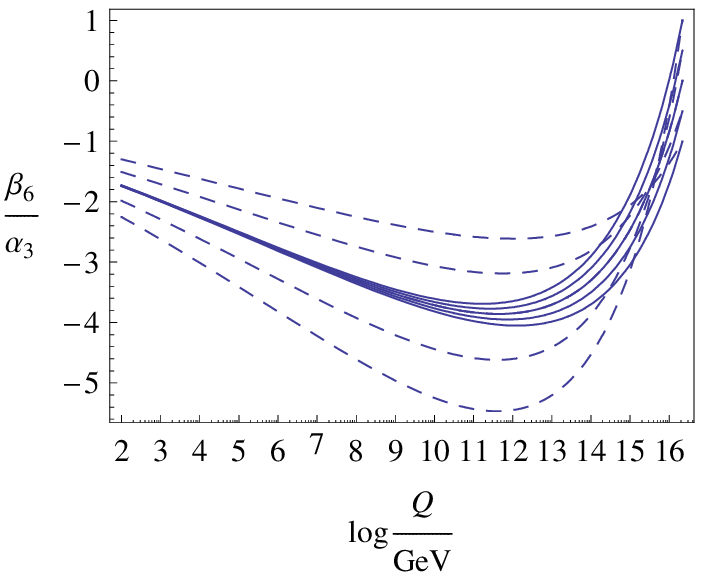}}
 \caption{The running of $\frac{\beta_6}{\alpha_3}$. For the solid curve only $\beta_6$ is non-zero at the high scale while for the dashed curve all $\beta_i$ are switched on.}
 \label{fig:b_5/a_2}
\end{figure}

To illustrate the behavior under the renormalization group, we analyze the
situation at several benchmark points. 
These points were chosen to be the so-called SPS points \cite{Allanach:2002nj}
(cf.\ table~\ref{tab:SPS}) amended by corrections in the MFV form. 
\begin{table}[h]
\centerline{\begin{tabular}{ccccc}
  Point & $m_0$ & $m_{\nicefrac{1}{2}}$ & $A$ & $\tan\beta$ \\ \hline
1a & \unit[100]{GeV} & \unit[250]{GeV} & \unit[-100]{GeV} & 10 \\ 
1b & \unit[200]{GeV} & \unit[400]{GeV} & 0 & 30 \\
2 & \unit[1450]{GeV} & \unit[300]{GeV} & 0 & 10\\
3 & \unit[90]{GeV} & \unit[400]{GeV} & 0 & 10\\
4 & \unit[400]{GeV} & \unit[300]{GeV} & 0 & 10 \\
5 & \unit[150]{GeV} & \unit[300]{GeV} & \unit[-1000]{GeV} & 5 \\ \hline
\end{tabular}}
\caption{Survey of SPS points.}
\label{tab:SPS}
\end{table}
Examples for the RG behavior are displayed in figures~\ref{fig:b_1/a_1}
and \ref{fig:b_5/a_2}. We show the ratio $\beta_1/\alpha_1$ and
$\beta_6/\alpha_3$, respectively. Note that these ratios coincide with $b_1/a_1$
and $b_6/a_3$ in the original MFV decomposition \cite{D'Ambrosio:2002ex}. These
ratios parametrize the deviations of the soft terms from unit
matrices.
In our illustrations, we use two different initial conditions for the
$\beta_i$. For the solid curve only the shown parameter is set non-zero at the
high scale, i.e.\ in \Figref{fig:b_1/a_1} only $\beta_1$ has a non-zero initial
value, while the dashed curves correspond to universal initial conditions for the
$\beta_i$. That is, we choose input values of the soft terms of the form
\eqref{eq:softmassdecomp} with 
\begin{equation}
 \alpha_i~=~m_0^2
 \quad\text{and}\quad
 \beta_j=0~\forall j\ne k\quad\text{or}\quad
 \beta_i~=~\text{universal}\;.
\end{equation}

We observe that the ratios get driven to non-trivial, i.e.\ non-zero, fixed
points. The corresponding low-energy fixed point values can be inferred from our
numerical approximations in appendix~\ref{app:Approximations}. 
These fixed points emerge from the competition from alignment and misalignment
processes, as discussed above.

\section{Beyond MFV}

The MFV ansatz is usually justified by a spurion argument. However, there are
some drawbacks to this reasoning. First of all, $G_\mathrm{flavor}$ (cf.\
equation~\eqref{eq:Gflavor}) is anomalous. Secondly, it is
hardly conceivable that one (spurion) field vev can give rise to a rank three
Yukawa coupling with hierarchical eigenvalues. From these considerations we
infer that the flavor symmetry will likely be broken by more than one field,
such that Yukawa couplings and corrections to soft parameters are proportional
to linear combinations of such fields with, in general, different coefficients.
In this picture one would expect corrections to the MFV scheme. 

Assuming that both new physics interactions and flavor models operate at the
high scale, it is worthwhile to  understand the implications of the presence of
non-MFV terms $\boldsymbol{\Delta  m}^2_f, \boldsymbol{\Delta  A}_f$, which we
add to the ansatz \eqref{eq:softmassdecomp}. That is, we decompose the soft terms
according to
\begin{subequations}\label{eq:nonMFV}
\begin{eqnarray}
 \boldsymbol{m}^2_f & = &  
 (\boldsymbol{m}^2_f)_\mathrm{MFV}  + \boldsymbol{\Delta m}^2_f\;,\\
 \boldsymbol{A}_f  & =  & 
 (\boldsymbol{A}_f)_\mathrm{MFV} + \boldsymbol{\Delta A}_f\;.
 \end{eqnarray}
\end{subequations}
We require the norm (cf.\ equation~\eqref{eq:norm}) of the non-MFV terms  to be
minimal, which makes the decomposition unambiguous. In other words, we demand
orthogonality between the MFV and the non-MFV terms, for instance
\begin{equation} 
\tr(\boldsymbol{\Delta m}^2_f)~=~0\;,\quad
 \tr(\Yu \Yu^\dagger\boldsymbol{\Delta m}^2_u)~=~0\;,\quad\text{etc.}
\end{equation}
As we know from our considerations in~\ref{ch:validity}, the non-MFV terms will
-- to high accuracy -- not be influenced by the running of the MFV terms. This
implies that, in a decomposition of the soft terms in MFV and non-MFV terms, the
evolution of non-MFV terms is governed by non-MFV terms only.  To make this
statement more precise, let us spell out the RGEs for the non-MFV terms. With
the additional requirement of orthogonality the derivation is analogous to the
one in section~\ref{ch:RGEs_MFV}. One obtains
\begin{subequations}
\begin{eqnarray}
16\pi^2\,\frac{\D}{\D t} \boldsymbol{\Delta m}_u^2
 &  =&  
2\boldsymbol{\Delta m}_u^2\,\Yu \Yu^\dagger
+4\,\Yu\,\boldsymbol{\Delta m}_Q^2\,\Yu^\dagger
+2\,\Yu \Yu^\dagger\,\boldsymbol{\Delta m}_u^2
+4\,\boldsymbol{\Delta}(\boldsymbol{A}_u\boldsymbol{A}_u^\dagger)\;,
\\
16\pi^2\,\frac{\D}{\D t} \boldsymbol{\Delta m}_d^2
& =&  
2\boldsymbol{\Delta m}_d^2\,\Yd \Yd^\dagger
+4\,\Yd\,\boldsymbol{\Delta m}_Q^2\,\Yd^\dagger
+2\,\Yd \Yd^\dagger\,\boldsymbol{\Delta m}_d^2
+4\,\boldsymbol{\Delta}(\boldsymbol{A}_d\boldsymbol{A}_d^\dagger)\;,
\\
16\pi^2\,\frac{\D}{\D t} \boldsymbol{\Delta m}_Q^2
 &=&  
\boldsymbol{\Delta m}_Q^2\,\left(
\Yu^\dagger \Yu+
\Yd^\dagger \Yd
\right)
+ \left(
\Yu^\dagger \Yu+
\Yd^\dagger \Yd
\right)\,\boldsymbol{\Delta m}_Q^2
\nonumber\\
& & {}
+2\,\Yu^\dagger\,\boldsymbol{\Delta m}_u^2\,\Yu
+2\,\Yd^\dagger\,\boldsymbol{\Delta m}_d^2\,\Yd
+2\,\boldsymbol{\Delta}(\boldsymbol{A}_u^\dagger\boldsymbol{A}_u)
+2\,\boldsymbol{\Delta}(\boldsymbol{A}_d^\dagger\boldsymbol{A}_d)\;,
\\
 16\pi^2\,
 \frac{\D }{\D t}\boldsymbol{\Delta A}_u
 & = & 
 \boldsymbol{\Delta A}_u\,
 \left[3\tr(\Yu\Yu^\dagger)
 +5\,\Yu^\dagger \Yu
 +\Yd^\dagger \Yd
 -\frac{16}{3}g_3^2-3g_2^2-\frac{13}{15}g_1^2 \right]
 \nonumber\\
  & & {}+\Yu\,
 \left[
  4\,\Yu^\dagger\,\boldsymbol{\Delta A}_u
 + 2\,\Yd^\dagger\,\boldsymbol{\Delta A}_d\right]\;,\\
 16\pi^2\,
 \frac{\D }{\D t}\boldsymbol{\Delta A}_d
 & = & 
 \boldsymbol{\Delta A}_d\,
 \left[\tr(3\,\Yd\Yd^\dagger
 	+\Ye^\dagger\Ye)
 +5\,\Yd^\dagger \Yd
 +\Yu^\dagger \Yu
 -\frac{16}{3}g_3^2-3g_2^2-\frac{7}{15}g_1^2 \right]
 \nonumber\\
 & & {}+\Yd\,
 \left[
  4\,\Yd^\dagger\,\boldsymbol{\Delta A}_d
 + 2\,\Yu^\dagger\,\boldsymbol{\Delta A}_u\right]\;,
 \label{eq:RGE4DelataAu}
\end{eqnarray}
\end{subequations}
where
$\boldsymbol{\Delta}(\boldsymbol{A}_f\boldsymbol{A}_f^\dagger)=\boldsymbol{\Delta
A}_f\,\boldsymbol{A}_f^\dagger+\boldsymbol{A}_f\,\boldsymbol{\Delta
A}_f^\dagger+\boldsymbol{\Delta A}_f\,\boldsymbol{\Delta A}_f^\dagger$. An
important point to notice is that the $\boldsymbol{\Delta m}^2_f$ terms get only
contributions from the Yukawas but not from the gauge couplings. This is not
true for the $\boldsymbol{\Delta A}_f$ terms, where the running is substantial.
However, the $\boldsymbol{\Delta A}_f$ terms cannot be too large since they are
constrained by FCNC processes and the requirement of avoiding of charge and color
braking minima \cite{Gabbiani:1996hi,Casas:1996de}. We have also checked that,
due to the hierarchical structure of the Yukawas, non-MFV terms will to a good
accuracy  stay non-MFV, i.e.\ orthogonal to the MFV terms, under the RGE. This
statement applies as long as the corrections $\boldsymbol{\Delta A}_f$ are not
too large, which we assume, as discussed. We would like to close by summarizing
the following observations:
\begin{enumerate}
\item
For vanishing $\boldsymbol{\Delta A}_f$ the $\beta$-functions of the $(\boldsymbol{ \Delta m}^2_f)_{\mathrm{off-diagonal}}$ are only proportional to the Yukawa
couplings. Hence the $\boldsymbol{(\Delta m}^2_f)_{\mathrm{off-diagonal}}$ stay almost constant.

\item
By contrast, the $\boldsymbol{\Delta A}_f$ do change due to the running. The
dominant contributions are a scaling effect proportional to the gauge couplings
and a lowering proportional to the top Yukawa. The net evolution can 
be approximated by $\boldsymbol{\Delta A}_f|_\mathrm{low-scale} \approx
(1\!-\!3)\cdot \boldsymbol{\Delta A}_f|_\mathrm{high-scale}$.
\end{enumerate}

\section{Discussion}

We have studied the scale-dependence of the structure of the MSSM soft masses
within the MFV framework. We find that, if the soft masses comply with the MFV
ansatz at one scale (such as $M_\mathrm{GUT}$), they can always be accurately
described in the MFV expansion. This implies that the RG evolution of soft terms
can then be expressed through the running (scalar) expansion parameters
$\beta_i$ (and $\alpha_i$). We have further studied the RG behavior of these
coefficients, and find that they get driven to  non-trivial fixed
points; i.e.\
that the low-energy values of $\beta_i$ are rather insensitive to their
`input' values at high energies. This has two important implications. First,
there is a degeneracy of parameters: regardless of what one assumes for the
$\beta_i$ parameters at the high scale one always obtains a very similar
phenomenology. Second, our results indicate that it might not be necessary to
keep the $\beta_i$ arbitrary if one works in the MFV scheme. Rather, for given
mSUGRA parameters the $\beta_i$ turn out to be restricted to very narrow ranges.
That is, if one takes the picture of the SUSY desert seriously and believes that
flavor originates from physics at high energies, there are in the MFV framework
only narrow ranges of parameters that need to be studied, at least as long the
ratio between scalar and gaugino masses is order unity.

We have also discussed corrections that go beyond the MFV decomposition.
It turns out that, in first approximation, in the case of the scalar
masses, non-MFV terms stay close to their boundary values. By contrast, in the
case of the trilinear couplings, non-MFV terms receive important corrections.

It is clear that our results can be extended in various respects. It should be
interesting to carry out an analogous analysis for the lepton sector. However,
due to the absence of gluino contributions, one might not expect a fixed point
behaviour which is as pronounced as in the quark sector. We have concentrated in
our work on moderate values of the Higgs vev ratio $\tan\beta$; extensions to
other, in particular large, values of $\tan\beta$ appear desirable. We have also
neglected phases in our presentation, to study their impact will be another
interesting task.

\subsection*{Acknowledgments}

We would like to thank  W.~Altmannshofer, A.~Buras, D.~Guadagnoli,
M.~Schmaltz, D.~Straub and M.~Wick for interesting discussions, and
B.~Allanach for correspondence. Two of us (P.P.~and~M.R.) would like to thank
the Aspen Center for Physics, where some this discussion was partially
initiated. This research is supported by the DFG cluster of excellence Origin
and Structure of the Universe, the Graduiertenkolleg "Particle Physics at the
Energy Frontier of New Phenomena" and the SFB-Transregios 27 "Neutrinos and
Beyond" by the Deutsche Forschungsgemeinschaft (DFG).

\clearpage
\appendix
\section{Numerical checks}
\label{app:MFVcheck}

In this appendix we describe how we numerically check the scale-independent
validity of the MFV decomposition.
For our numerical calculations we use SOFTSUSY Version 2.0.14
\cite{Allanach:2001kg}. We restrict ourselves to real matrices only (partially
because of the corresponding limitation of SOFTSUSY). Apart from the usual
GUT-relations
\begin{eqnarray}
 M_1 &= & M_2 ~=~ M_3 ~=:~ m_{\nicefrac{1}{2}} \;, \nonumber\\
 \alpha_1 &=& \alpha_2 ~=~ \alpha_3 ~=~ m^2_{H_u} ~=~ m^2_{H_d} ~=:~ m_0^2,
 \quad \boldsymbol{m}^2_e ~=~ \boldsymbol{m}^2_L ~=~ m_0^2\,\mathds{1}\;,\nonumber\\
 \alpha_4 &=& \alpha_5 ~=:~ A\;,
\label{eq:universal}
\end{eqnarray}
we consider here only universal $\beta_i$:
\begin{equation}
 \beta_1~=~\dots~=~\beta_6~=:~b\,m_0^2\;,\quad
 \beta_7~=~\beta_8~=~b\, A\;.
\end{equation}  
Restricting ourselves to $\tan\beta=10$, we perform a scan over the following
region in parameter space:
\begin{gather*}
 \unit[-1000]{GeV}~<~ A~ <~ \unit[1000]{GeV}\;, \qquad |A| ~\leq~ m_0\;, \quad
 \unit[200]{GeV}~ <~ m_{\nicefrac{1}{2}} ~<~ \unit[500]{GeV}\;, \\
 \unit[100]{GeV} ~<~ m_0 ~<~ \unit[1500]{GeV}\;,\quad
 |\beta_{1,2,3,4,5,6}| ~\leq ~m_0^2\;, \qquad |\beta_{7,8}|~ \leq ~|A|\;.
\end{gather*}

At the low scale, defined by SOFTSUSY as $\sqrt{m_{\tilde t_1}\, m_{\tilde
t_2}}$, a best fit decomposition is used to minimize the absolute difference to
the form of equation~\eqref{eq:softmassdecomp}. To this end, we use the matrix
norm
\begin{equation}
 |M_{ij}|~=~\sqrt{ \sum_{ij} |M_{ij}|^2}\;.
 \label{eq:norm}
\end{equation}
Define now $\boldsymbol{\Delta m}_f$ $\left(\boldsymbol{\Delta A}_f\right)$ as the
part of $\boldsymbol{m}_f$ ($\boldsymbol{A}_f$) which is orthogonal to the MFV
decomposition at the low scale (cf.\ equation~\eqref{eq:nonMFV}). Then for all the points in our scan the ratio
$\frac{|\boldsymbol{\Delta m}_f|}{|\boldsymbol{m}_f|}$ 
($\frac{|\boldsymbol{\Delta A}_f|}{m_{\nicefrac{1}{2}}}$) lies below the indicated number (table~\ref{tab:Deviations}). 
We normalize $\boldsymbol{\Delta A}_f$ to $m_{\nicefrac{1}{2}}$ rather than
$|\boldsymbol{A}_f|$ since the later can approach zero at the low scale. Notice
also that we truncate the MFV decomposition as specified in 
\eqref{eq:softmassdecomp}. The deviations (table~\ref{tab:Deviations}) are of
the order of higher-order MFV terms, i.e.\ we expect that the MFV approximation
gets practically perfect when higher order terms are included.

\begin{table}[h!]
\centerline{\begin{tabular}{l|ccccc}
 quantity &
 $\boldsymbol{m}^2_Q$ & $\boldsymbol{m}^2_u$ & $\boldsymbol{m}^2_d$ &
 $\boldsymbol{A}_u$ & $\boldsymbol{A}_d$ \\ \hline
 deviation &
 $10^{-7}$ & $10^{-6}$ & $10^{-5}$ & $10^{-2}$ &  $10^{-3}$ \\
\end{tabular}}
\caption{Deviations from scale independency.}
\label{tab:Deviations}
\end{table}

We observe that for the bilinear soft masses the MFV-decomposition holds with
great accuracy. In the case of the trilinears the error is of the order
$\frac{m_c}{m_t}\approx 1\%$, as one might have expected.

In summary, for $|\beta_i| \leq \alpha_j$ (comparing only coefficients of same
mass dimension), soft terms which are in the MFV form at the GUT scale will be
in this form at the low scale with good precision.

\begin{landscape}
\section{Approximations on low-energy MFV coefficients}
\label{app:Approximations}

We numerically solve the RGEs as described in appendix~\ref{app:MFVcheck}, but
with only one $\beta_i$ set different from zero.

The following formulae reproduce the exact SOFTSUSY results up to an error of
\begin{gather*}
 \frac{|\alpha_{i,\textrm{fit}}-\alpha_{i,\textrm{SOFTSUSY}}|}{  |\alpha_{i,\textrm{SOFTSUSY}}|} < 0.1 \quad \textrm{and} \\
 \frac{|\beta_{1,2,3;4;5,\textrm{fit}}-\beta_{1,2,3;4;5,\textrm{SOFTSUSY}}|}{|\alpha_{1;2;3,\textrm{SOFTSUSY}}|},
\frac{|\beta_{7,8,\textrm{fit}}-\beta_{7,8,\textrm{SOFTSUSY}}|}{m_{\nicefrac{1}{2}}} < 0.02\;.
\end{gather*}

In the following formulae, the variables on the left hand side denote the values
at the low scale, while those on the right hand side are the quantities at the
high scale. 
\tabcolsep=0.5pt 
\begin{center}
\begin{tabular}{rclllllll}
$\alpha_1$ &~=~& $+0.94\,m_0^2$&$+5.04\,m^2_{\nicefrac{1}{2}}$\\ 
$\alpha_2$ & = & $+0.95\,m_0^2$&$+4.72\,m^2_{\nicefrac{1}{2}}$\\ 
$\alpha_3$ & = & $+0.95\,m_0^2$&$+4.61\,m^2_{\nicefrac{1}{2}}$\\
$\alpha_4$ & = && $-2.00\,m_{\nicefrac{1}{2}}$&& $+0.32\,A$ \\ 
$\alpha_5$ & = && $-3.23\,m_{\nicefrac{1}{2}}$&& $+0.98\,A$   \\
$\beta_1$ & = & $-0.41\,m_0^2$   & $-0.96\,m^2_{\nicefrac{1}{2}}$ & $+0.16\,A\,m_{\nicefrac{1}{2}}$ & $-0.04\,A^2$ &  $+0.27\,\beta_1-0.03\,\beta_5$\\ 
$\beta_2$ & = & $-0.43\,m_0^2$   & $-1.38\,m^2_{\nicefrac{1}{2}}$  & $+0.57\,A\,m_{\nicefrac{1}{2}}$ & $-0.15\,A^2$  &  $-0.02\,\beta_1+0.1\,\beta_2+0.01\,\beta_5$\\
$\beta_3$ & = && $+0.13\,m^2_{\nicefrac{1}{2}}$  & $-0.13\,A\,m_{\nicefrac{1}{2}}$&  $+0.04\,A^2$ & $+0.02\,\beta_1+0.03\,\beta_3-0.01\beta_5-(0.01\,\beta_7+0.04\,\beta_8)\,A+(0.03\,\beta_7+0.08\,\,\beta_8)\,m_{\nicefrac{1}{2}}$  \\
$\beta_5$ & = & $-0.83\,m_0^2$  &  $-1.96\,m^2_{\nicefrac{1}{2}}$  & $+0.32\,A\,m_{\nicefrac{1}{2}}$ & $-0.09\,A^2$ &  $-0.07\,\beta_1+0.24\,\beta_5$ \\
$\beta_6$ & = & $-0.86\,m_0^2$  & $-2.57\,m^2_{\nicefrac{1}{2}}$  & $+0.94\,A\,m_{\nicefrac{1}{2}}$ & $-0.25\,A^2$  &   $-0.07\,\beta_1+0.01\beta_5+0.12\beta_6-0.14\,A\,\beta_8+0.25\,m_{\nicefrac{1}{2}}\,\beta_8 $ \\
$\beta_7$ & = &&$+0.51\,m_{\nicefrac{1}{2}}$&& $-0.27\,A$ &$+0.10\,\beta_7$ \\
$\beta_8$ & = &&$+0.27\,m_{\nicefrac{1}{2}}$&& $-0.14\,A$ &$+0.30\,\beta_8$ \\
\end{tabular}
\end{center}
\end{landscape}


\begin{thebibliography}{10}

\bibitem{Dimopoulos:1981zb}
S.~Dimopoulos and H.~Georgi, Nucl. Phys. \textbf{B193} (1981), 150.

\bibitem{Chivukula:1987py}
R.~S. Chivukula and H.~Georgi, Phys. Lett. \textbf{B188} (1987), 99.

\bibitem{Buras:2000dm}
A.~J. Buras, P.~Gambino, M.~Gorbahn, S.~J{\"a}ger, and L.~Silvestrini, Phys.
  Lett. \textbf{B500} (2001), 161--167,  [hep-ph/0007085].

\bibitem{D'Ambrosio:2002ex}
G.~D'Ambrosio, G.~F. Giudice, G.~Isidori, and A.~Strumia, Nucl. Phys.
  \textbf{B645} (2002), 155--187,  [hep-ph/0207036].

\bibitem{Martin:1993zk}
S.~P. Martin and M.~T. Vaughn, Phys. Rev. \textbf{D50} (1994), 2282,
  [hep-ph/9311340].

\bibitem{Brignole:1993dj}
A.~Brignole, L.~E. Ib{\'a}{\~n}ez, and C.~Mu{\~n}oz, Nucl. Phys. \textbf{B422}
  (1994), 125--171,  [hep-ph/9308271].

\bibitem{Choudhury:1994pn}
D.~Choudhury, F.~Eberlein, A.~K{\"o}nig, J.~Louis, and S.~Pokorski, Phys. Lett.
  \textbf{B342} (1995), 180--188,  [hep-ph/9408275].

\bibitem{Brax:1995up}
P.~Brax and C.~A. Savoy, Nucl. Phys. \textbf{B447} (1995), 227--251,
  [hep-ph/9503306].

\bibitem{Chankowski:2005jh}
P.~H. Chankowski, O.~Lebedev, and S.~Pokorski, Nucl. Phys. \textbf{B717}
  (2005), 190--222,  [hep-ph/0502076].

\bibitem{Allanach:2002nj}
B.~C. Allanach et~al.,  (2002),  hep-ph/0202233.

\bibitem{Gabbiani:1996hi}
F.~Gabbiani, E.~Gabrielli, A.~Masiero, and L.~Silvestrini, Nucl. Phys.
  \textbf{B477} (1996), 321--352,  [hep-ph/9604387].

\bibitem{Casas:1996de}
J.~A. Casas and S.~Dimopoulos, Phys. Lett. \textbf{B387} (1996), 107--112,
  [hep-ph/9606237].

\bibitem{Allanach:2001kg}
B.~C. Allanach, Comput. Phys. Commun. \textbf{143} (2002), 305--331,
  [hep-ph/0104145].

\end{thebibliography}

\providecommand{\bysame}{\leavevmode\hbox to3em{\hrulefill}\thinspace}

\end{document}